\documentclass[a4paper]{aipproc}
\layoutstyle{6x9}
\usepackage{epsfig}
\def\G1915{GRS $1915$+$105$}
\def\X1550{XTE J$1550$-$564$}
\def\J1655{GRO J$1655$-$40$}
\def\eg{{\it e.g.} }
\def\etal{{\em et al. } }

\def\dd #1 {{\frac{\partial}{\partial #1}}}

\def\cs2{c_{S}^2}

\def\ltsima{$\; \buildrel < \over \sim \;$}
\def\simlt{\lower.5ex\hbox{\ltsima}}
\def\gtsima{$\;\buildrel>\over\sim\;$}
\def\simgt{\lower.5ex\hbox{\gtsima}}

\begin{document}

\author{J. Rodriguez}{
	address={CEA/DSM/DAPNIA/Service d'Astrophysique 
	(CNRS URA 2052), 91191 Gif Sur Yvette Cedex, France},
	email={jrodriguez@cea.fr},
}
\author{Ph. Durouchoux}{
	address={CEA/DSM/DAPNIA/Service d'Astrophysique 
	(CNRS URA 2052), 91191 Gif Sur Yvette Cedex, France},
}
\author{M. Tagger}{
	address={CEA/DSM/DAPNIA/Service d'Astrophysique 
	(CNRS URA 2052), 91191 Gif Sur Yvette Cedex, France},
}
\title{RXTE Observations of GRS 1915+105}
\begin{abstract}

We analyse a set of three RXTE Target of Opportunity observations of the
Galactic microquasar \G1915, observed in April 2000.  We concentrate on
the timing properties of the source, and examine the properties of a low
frequency QPO, with its harmonic, in several energy ranges.  The source
was found in two different states of the spectro/temporal classification
of Belloni \etal (2000), and exhibited in the three observations a
strong, low frequency QPO together with a strong harmonic.  We discuss
the properties of the QPO, of its harmonic and of their spectral
behaviour in the framework of the Accretion Ejection Instability (AEI)
(Tagger \& Pellat, 1999; Varni\`ere, Rodriguez \& Tagger, 2001;
Rodriguez \etal, 2001).
\end{abstract}

\maketitle

\section{Introduction : QPO's in Black Hole Binaries}
\begin{figure}[htpb]
\epsfig{file=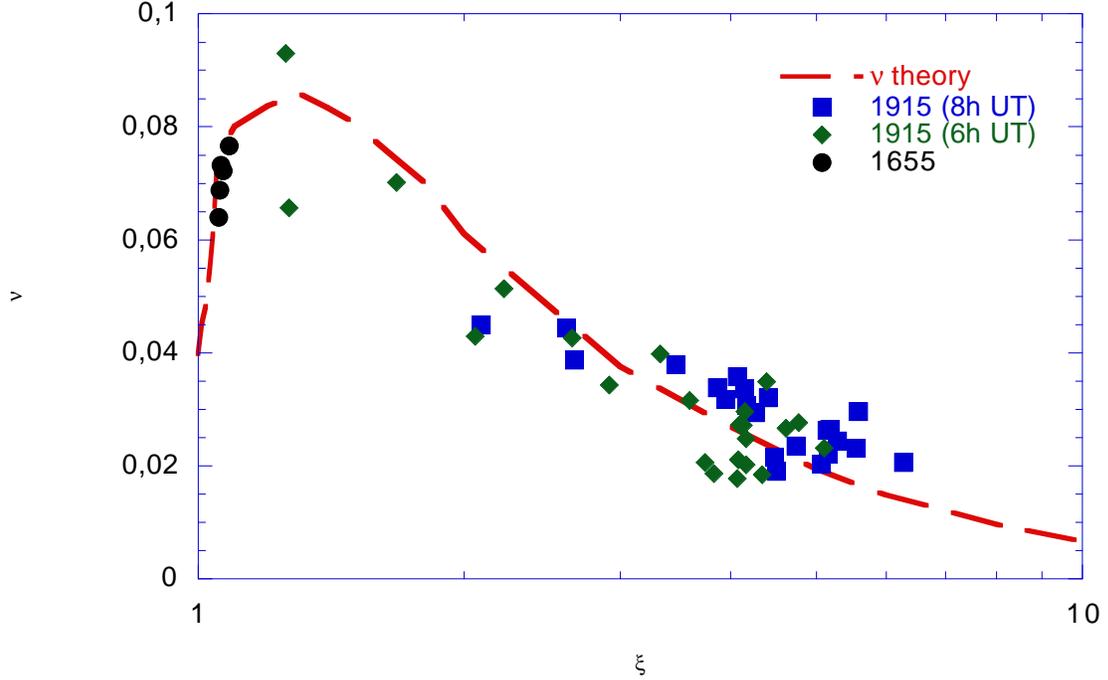,width=\columnwidth} \caption{Plot of the QPO
frequency $\nu_{QPO}$ vs.  the inner disk radius, $r_{int}$, for \G1915
and \J1655, together with the theorical curve; the X axis is in units of
$\xi=r_{int}/r_{Last\ Stable\ Orbit}$, and the Y axis is in units of the
keplerian frequency at the last stable orbit.  The fit is obtained with
the estimated mass of \J1655, while for \G1915 it gives a mass of $\sim
20 M_{\odot}$, compatible with independent studies.}
\label{fig:correl}
\end{figure}

X Ray binaries exhibit quasi-periodic behaviors on time scales ranging
from millisecond to days or more.  Systematic monitoring of these
sources, and the use of instruments with high timing resolution such as
the Rossi X ray Timing Explorer (RXTE), it has become possible to
distinguish several type of QPO's, based on their frequency, and to
correlate their properties with the spectral state of the source.\\
In the case of \G1915 at least three types of QPO's have so far been
detected : a 67Hz one during soft high states (Morgan, Remillard \&
Greiner, 1997), a 67mHz one, also present during soft high states
(Morgan, Remillard \& Greiner, 1997), and a $1-10$Hz variable QPO which
appears to be ubiquitous during the low hard state.  We focus here
only on this ubiquitous QPO, for which several studies have pointed out
correlations between the spectral and the temporal parameters, such as the
source flux and QPO frequency (Markwardt \etal, 1999), or the inner disk
temperature and QPO frequency (Muno \etal, 1999).\\
An intriguing result was recently reported by Sobczak \etal (1999), who
found that in \J1655 the correlation between the QPO frequency and the
disk inner radius was the opposite of that found in \X1550.  In previous
work we have confirmed this reversed correlation and compared \J1655
with \G1915 (Varni\`ere, Rodriguez \& Tagger, 2001; Rodriguez \etal,
2001).  We showed that it could be explained (see fig \ref{fig:correl})
if the QPO is identified with the Accretion-Ejection Instability (AEI)
of Tagger \& Pellat (1999).
\section{The source on April $17^{th}$}
\begin{figure}[htpb]
\epsfig{file=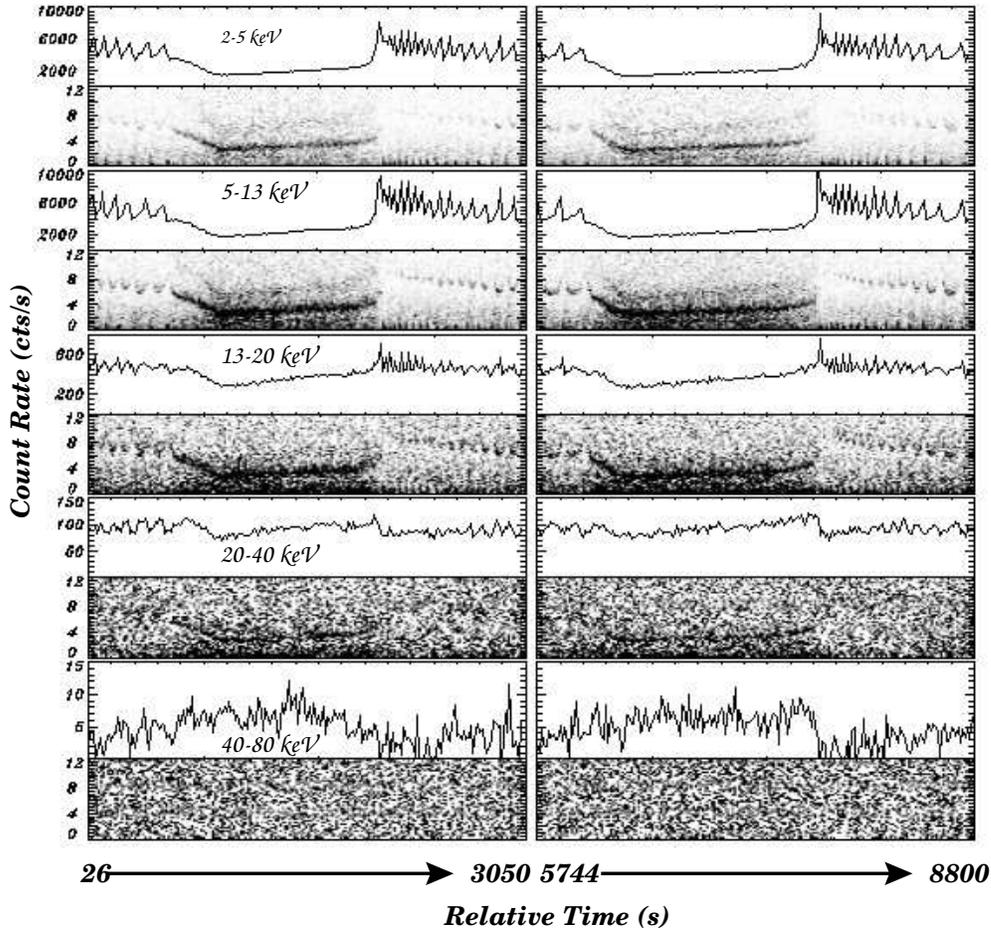,width=\columnwidth} \caption{Dynamical
power spectra of the source on April $17^{th}$, in five PCA energy
ranges.  The X axis is in seconds since the beginning of the time
intervals studied.  The Y axis is in count/sec.  in the upper panels,
and in Hz in the lower ones.  The gap in the data corresponds to
occultations during the orbit}
\label{fig:17-04qpo}
\end{figure}
We show on figure \ref{fig:17-04qpo} dynamical power spectra of the
source in five PCA energy bands.  The source is in the $\nu$ state of
Belloni \etal (1999), characterized by a $\sim 30$ mn cycle between a
high/soft and a low/hard state.  Each interval starts with the source in
the high/soft state, showing episodically the QPO. After the ``dip'' (at
$t\simeq 600 s.$ and $6000 s.$) marking the transition to the low/hard
state, the QPO is stronger and shows a prominent harmonic, appearing as
a second dark lane in the lower panels of figure 2.  A strong X-ray
spike, at times $t \simeq 2000 s.$ and $7700 s.$, marks a return to the
high state, with again episodic occurence of the QPO; incidentally the
fundamental frequency of the QPO is close to that of the harmonic just
before the spike.  \\
Comparing the lightcurves to the variations of the QPO frequency, we
find that although the QPO power is stronger in the 5-13 keV band, its
frequency variation is better correlated with the softer flux; no QPO is
observed above 40 keV; we note that the large dips are smoothed while
rising in energy, so that the soft spike corresponds to a sudden
decrease of the hard emission; this can be interpreted as the
disappearance of a part of the corona, blown away by a sudden ejection
coincident with the spike, as already seen during multiwavelength
observations of similar states (Eikenberry \etal, 1998; Mirabel \etal,
1998).
\section{April $22^{nd}$ and $23^{rd}$}
\begin{figure}[htpb]
\epsfig{file=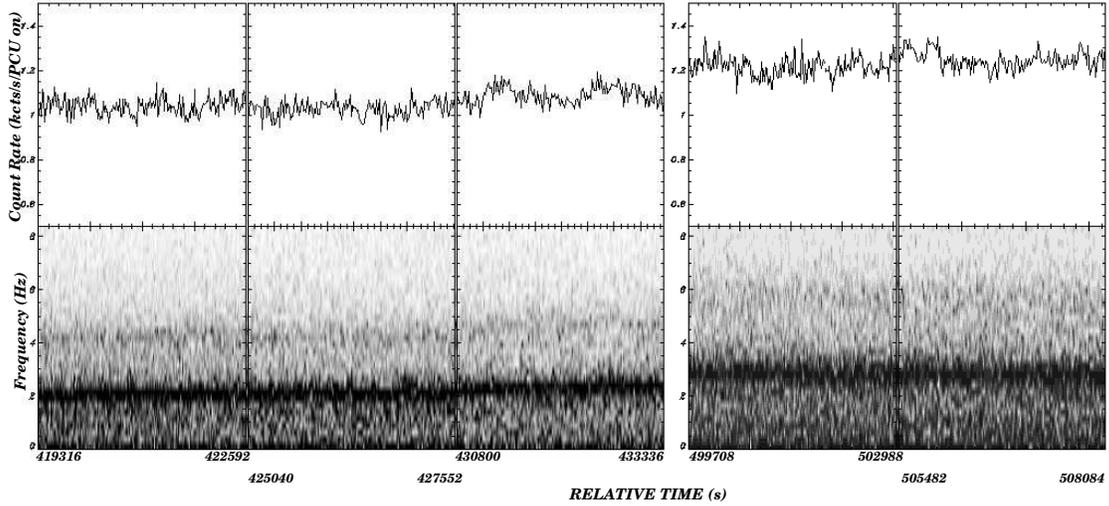,width=\columnwidth}
\caption{Standard lightcurves (upper pannels), and dynamical power
spectra of the source (lower pannels) on April $22^{nd}$and $23^{rd}$.
X axis is in second, time 0 is the same as in figure
\ref{fig:17-04qpo}.}
\label{fig:22-23_qpo}
\end{figure}
\begin{figure}[htpb]
\epsfig{file=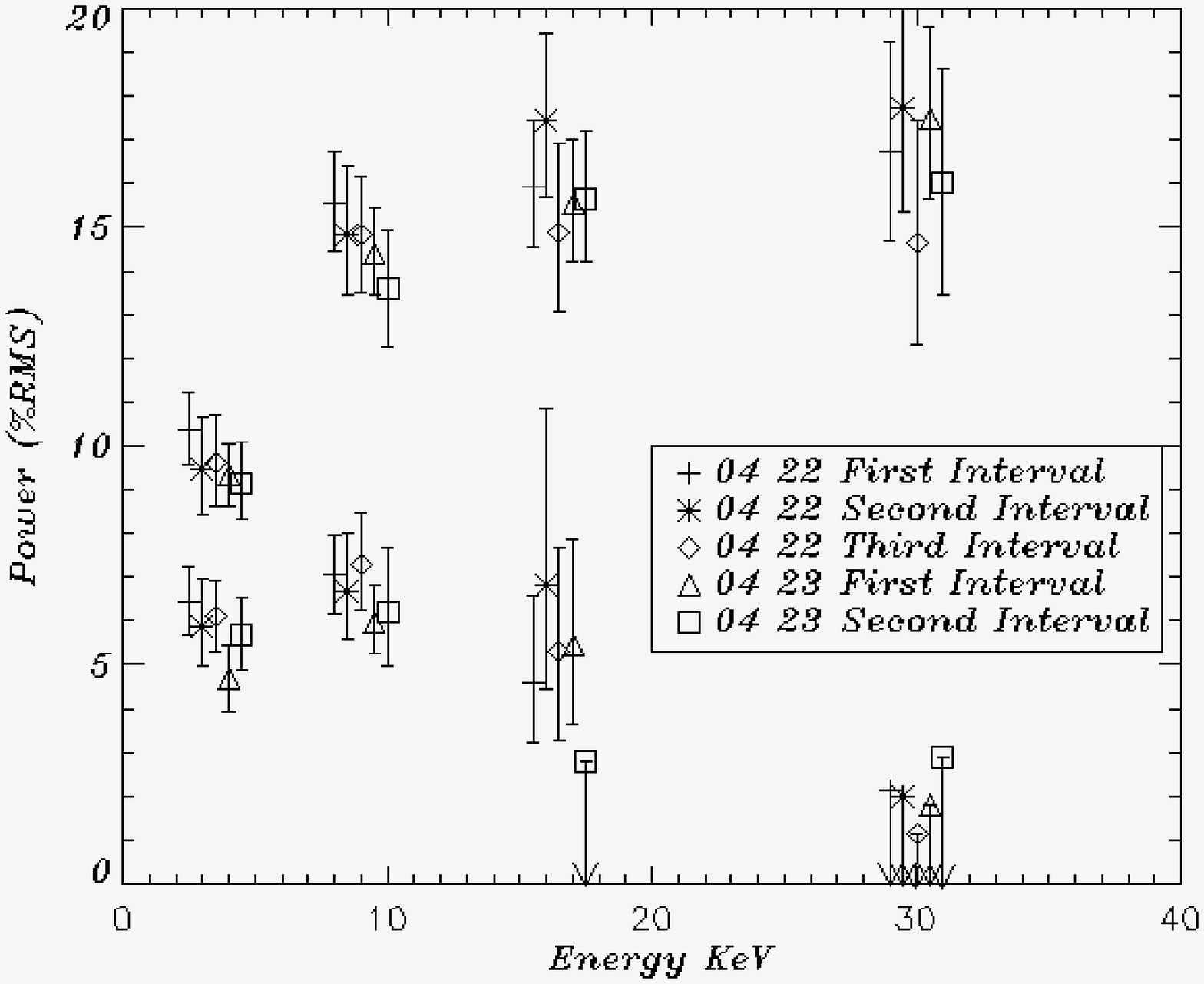,width=\columnwidth} \caption{Plot of the
power(in unit of \% RMS) vs.  Energy range (keV), for the QPO and its
harmonic, in four bands defined in the text; the higher range is not
represented since the lack of flux above 40 keV did not permit a
detection of a QPO.  The lower points correspond to the harmonic, the
upper ones to the fundamental.}
\label{fig:qpovsenergy}
\end{figure}

The source is in a low and steady state at these dates, and as we could
expect from previous studies (Markwardt \etal, 1999; Muno \etal, 1999;
Rodriguez \etal, 2001) the QPO frequency is fairly constant (figure
\ref{fig:22-23_qpo}), over the whole observations, with a value of $\sim
2.15$Hz on  the first two  intervals of April 22, $\sim 2.37$Hz 
during the third
one, and $\sim 2.9$Hz on April 23.\\
We again extracted lightcurves and produced power spectra in 5 PCA
energy channels ($\sim 2-5$keV,$\sim 5-13$keV, $\sim 13-20$keV,$\sim
20-40$keV, and above 40 keV).  power vs.  energy, for the QPO and its
harmonic, are plotted in figure \ref{fig:qpovsenergy}.
Only four ranges are plotted since the last one suffers from the lack of
flux, which did not allow us to extract QPO parameters.  Nevertheless,
we find that the QPO power increases with the energy, up to about 30
keV, whereas the harmonic decreases above 10 keV. These distinct
spectral behaviour thus represent a new challenge for models of the QPO.
\section{Conclusion}
Our study confirms and expands the conclusion (Markwardt \etal, 1999;
Muno \etal, 1999) that the QPO frequency is better correlated with the
softer flux, which tends to show that it has its origin in the disk,
but that it affects more strongly the flux at higher energies usually
considered to be emitted by the corona. In addition figure
\ref{fig:17-04qpo} shows that the presence of the harmonic is correlated
to a strong coronal emission; whenever the latter disapear, only the
fundamental remains.\\
Tagger \& Pellat (1999) have shown that the inner region of the disk
could exhibit what they called an Accretion-Ejection instability,
extracting energy and angular momentum from the disk and transferring
them to Alfven waves emitted toward the corona.  The instability forms a
steady spiral pattern, rotating at a frequency compatible with that of
the ``ubiquitous'' QPO (see Varni\`ere and Tagger, these proceedings).
In this context, the harmonic can be seen as a diagnostic of the
non-linear development of the instability, \eg the formation of a hot
point or a spiral shock in the disk (Rodriguez \etal, 2001). A
useful analogy can be made with the gaseous shock marking the spiral
arms of galaxies. In this
context the diappearance of the harmonic during the high state could
correspond to a weaker instability leading to less sharp non-linear
features; the decrease in the power of the harmonic at high energies,
contrasting with the behavior of the fundamental, would indicate that
the harmonic does not propagate to the corona.

\end{document}